# Online *in-situ* X-ray diffraction setup for structural modification studies during swift heavy ion irradiation

C. Grygiel*, H. Lebius, S. Bouffard, A. Quentin, J. M. Ramillon, T. Madi, S. Guillous, T. Been, P. Guinement, D. Lelièvre, and I. Monnet

*CIMAP, CEA – CNRS – ENSICAEN – UCBN, BP 5133, 14070 Caen Cedex 5, France*

The high energy density of electronic excitations due to the impact of swift heavy ions can induce structural modifications in materials. We present a X-ray diffractometer called ALIX, which has been set up at the low-energy IRRSUD beamline of the GANIL facility, to allow the study of structural modification kinetics as a function of the ion fluence. The X-ray setup has been modified and optimized to enable irradiation by swift heavy ions simultaneously to X-ray pattern recording. We present the capability of ALIX to perform *simultaneous irradiation - diffraction* by using energy discrimination between X-rays from diffraction and from ion-target interaction. To illustrate its potential, results of *sequential* or *simultaneous irradiation - diffraction* are presented in this article to show radiation effects on the structural properties of ceramics. Phase transition kinetics have been studied during xenon ion irradiation of polycrystalline MgO and $SrTiO_3$. We have observed that MgO oxide is radiation-resistant to high electronic excitations, contrary to the high sensitivity of $SrTiO_3$, which exhibits transition from the crystalline to the amorphous state during irradiation. By interpreting the amorphization kinetics of $SrTiO_3$, defect overlapping models are discussed as well as latent track characteristics. Together with a transmission electron microscopy study, we conclude that a single impact model describes the phase transition mechanism.

## I. Introduction

Materials under irradiation of swift heavy ions can undergo structural evolutions induced by high electronic excitation which leading to modifications of their macroscopic properties.[1,2] To foresee the consequences of irradiation in radiative environments (nuclear or space) or to control the evolution of material properties induced by irradiation (doping of semiconductor, hardening of polymers, nanostructuration), it is necessary to have a full description of processes involved during ion irradiation (from ion/target interaction to material relaxation) as well as effects on physico-chemical properties.[3-5] After years of research on the properties of irradiated materials, numerous questions are still unsolved. One of the most important questions is the determination of the mechanisms governing the phase transition induced by irradiation of ceramics. The study of these kinetics as a function of the ion fluence has been historically carried out by using different samples irradiated at each individual fluence.[6] Alignment variations and differences between individual samples are a major error source, thus the *ex-situ* experiments are limited in their accuracy. Additionally, more irradiation beam time is needed as the cumulative fluence is larger than the maximum fluence. This maximum fluence is sufficient when only one target is used. *In-situ* experiments allow time-saving and avoid reproducibility problems. In the framework of the publication we call *in-situ* for experiments where the targets do not leave the vacuum chamber, therefore it is preferable to have a fixed target position, and then the observed volume/surface and the irradiated volume coincide. For this reason, *in-situ* X-ray diffraction (XRD) has been used at ion-beam accelerator facilities in order to study structural evolutions, phase transitions and amorphization, induced by swift heavy ion irradiation. At the middle-energy beamline (SME) of Grand Accélérateur National d'Ions Lourds (GANIL), the setup called "CHEXPIR" was developed some years ago where the targets are at room temperature, the detection angle spans from 10° to 60° (2θ) using a monochromatized radiation source.[7] A 4-circle diffractometer optimized for single crystal studies is mounted at the M2-branch of the Universal Linear Ion Accelerator (UNILAC) at GSI facility.[8, 9] At the low-energy beamline (IRRSUD) of GANIL a new X-ray diffractometer (called "ALIX"), upgraded of CHEXPIR, has been recently set up in order to make available *in-situ* X-ray diffraction simultaneously to irradiation. *Simultaneous* refers to *in-situ* irradiation at the same time as online analysis (i.e. diffraction). *Sequential* refers to irradiation and analysis performed at different times, as it is done on the other equipments (CHEXPIR or M2-GSI). An *in-situ* XRD setup is mounted at Inter-University Accelerator Centre (IUAC) using variable temperatures, to our knowledge incident as well as diffracted X-ray angles are geometrically limited and no simultaneous *in-situ* experiments are available.[10]

---

* Corresponding author, electronic mail: *grygiel@ganil.fr*

As an example, an insulating ionic-covalent material $SrTiO_3$, from *$ABO_3$* perovskite family, has been selected for structural study with ALIX setup under swift heavy ion irradiation. The $SrTiO_3$ perovskite is a widely used material due to its structural and physical properties and it is now considered as an ideal substrate for epitaxial growth of superconducting or magnetic thin films.[11] Moreover, the large tolerance factor of perovskite allows numerous substitutions. Perovskites with actinides on the *A* site have been shown as potential candidates for actinide immobilization in nuclear waste.[12] Therefore, $SrTiO_3$ represents a model perovskite for which it is important to know its stability under irradiation conditions. Studies have already shown its behaviour under low-energy (nuclear stopping power regime) ion irradiation showing an amorphization due to disorder accumulation[13, 14] and a recrystallization has been observed at the amorphous-crystalline interface by annealing.[14-16] Until now, damages due to high energy ion irradiation have been less studied, amorphization has been observed above an electronic stopping power threshold of 12 keV/nm,[17, 18] however its reaction kinetics remains an open question and will be discussed in this article.

We will detail the design decisions of ALIX, as well as modifications required to set it up on the IRRSUD beamline. Calibration measurements using reference $LaB_6$ powder will be shown. Simultaneous irradiation and diffraction requires energy discrimination of the diffracted X-rays. This mode has been tested on MgO ceramics during irradiation. Finally results of radiation effects on structural properties of $SrTiO_3$ will be discussed. All irradiations have been performed with 92 MeV Xe ions.

## II. *In-situ* X-ray diffraction during swift heavy ion irradiation

### 1. Description of the *in-situ* XRD instrument setup "ALIX"

In this section the technical details of the equipment are discussed, especially consequences of the beamline choice and the X-ray diffraction requirements. The GANIL facility provides swift heavy ions in three different energy regimes. Ions from the high-energy beamline (HE, 24 – 95 MeV/u) as well as from the middle-energy beamline (SME, 4 – 14 MeV/u) have larger penetration depth than those of the X-rays used, but the ions generate sample activation and thus, a non-homogeneous increase of the background in XRD patterns. Therefore it is necessary to wait a long time before each XRD measurement in order to reduce the background.[19] At the low-energy beamline (IRRSUD, 0.3 to 1 MeV/u), no activation takes place, due to the energy, which is below the nuclear reaction threshold. However, at these energies the penetration depth of the ions is reduced to some micrometers, and the electronic stopping power varies strongly after the first 1-10 µm. In figure 1, the electronic stopping power ($S_e$, in keV/nm) as well as the displacements induced by nuclear collisions (in dpa, displacements per atom, for a fluence of $1 \times 10^{14}$ ions cm$^{-2}$) are plotted against the penetration depth of 92 MeV Xe ions in $SrTiO_3$. The values were calculated using the SRIM 2008 code assuming displacement energies of 25 eV for all atoms.[20] It is clearly seen that the electronic stopping power is the main contribution to the overall energy loss close to the surface and that it decreases continuously with increasing penetration depth. On the first µm the electronic energy loss changes by about 10%. Therefore, in order to study irradiation effects at a fixed energy loss value, it is necessary to limit the study to the topmost part of the sample. This can be achieved by using grazing incidence. Considering an incidence angle of the X-ray at 1° from the SrTiO$_3$ surface, the thickness probed is around 700 nm with a mean S$_e$ of 20 keV/nm.

A Bruker AXS D8 Discover diffractometer with θ/θ geometry with a Göbel mirror (parabolically-shaped multilayer mirror, 40 mm) was used. The parallel X-ray beam allows grazing incidence diffraction due to its low divergence, besides an enhancement of the X-ray beam intensity. The Cu emitter with the K$_{\alpha 1}$ line at 1.54056 Å was used. A 1D detector, Vantec-1, with a 10° window was chosen. This allows simultaneous diffraction measurements over a large diffraction angle range, therefore minimising the necessary beamtime at the facility. The detector has a resolution of 0.00626°, and allows quick scans with a good signal-to-background ratio and an effective energy discrimination between X-rays from diffraction and from ion-target interaction.

In order to adapt the diffractometer to the IRRSUD beamline, as shown in fig. 2, several modifications were done. Two apertures have been placed to align and later guide the ion beam. The sample holder has been fitted with tilt/shift movements in order to position the sample surface to coincide with the intersection between ions and X-rays. The ion beam hits the surface under 18° with respect to the surface normal. This allows a larger angle range of the detector as he can move below the ion beam. To get a homogeneous fluence on the sample surface, the ion beam is scanned over around 5 cm² which is much larger than the typical target surface of around 1 cm². A removable alumina plate is used to check the scanned beam area. The irradiation/diffraction chamber is kept at room temperature under high vacuum conditions with pressures in the 10$^{-6}$-10$^{-7}$ mbar region, and is equipped with 50 μm thick kapton windows for the passage of the X-ray beams. These windows allow incident X-ray angles from -5° to 10° and diffracted X-ray angles from -5° to 120° with respect to the surface normal. In the geometry described here, only detector angle (i.e. 2θ angle) is changing, whereas the X-ray source angle and sample are fixed. It is worth noting that conventional θ/2θ scans, where source and detector angle changes are coupled, are also possible during simultaneous *in-situ* experiments by using another kapton window of 190° aperture.

## 2. Setup validation: measurement of standard reference LaB$_6$ powder

In order to test the alignment of the goniometer and the target surface, using the above-mentioned tilt/shift mechanic, a validation has to be performed of the ALIX device with standard structure analysis. The conventional θ/2θ pattern has been recorded for a standard reference material, NIST SRM660b LaB$_6$ powder, generally used in calibration of diffraction line positions and line shapes. The XRD data, recorded in the 2θ range of 10-140° with a step size of 0.0313°, were analysed by Rietveld refinements with the FullProf software.[21] The Thompson-Cox-Hastings function was selected to refine the peak profile.[22] The observed, calculated, and difference profiles are plotted in figure 3. The refinement converges to agreement factors of $R_{wp}$=9.4%, $R_B$= 4.1% and GofF=1.17 ($R_{wp}$, $R_B$ are inferior to 10% and GofF tends to 1), assuming the theoretical cubic *Pm-3m* symmetry of LaB$_6$. These results show the good alignment of the modified sample holder. Therefore, reliable measurements of XRD patterns are assured.

**3. Energy discrimination: case of MgO polycrystalline pellet**

Another step for the device validation is to check if X-ray diffraction patterns can be recorded during ion irradiation. The material used for the validation should be insensitive to the electronic stopping power, therefore the structure should not change under irradiation. Polycrystalline MgO is an ionic oxide known to exhibit a low sensitivity to electronic excitation and to be relatively radiation-resistant to high ion fluence.[23] Patterns have been recorded at successive fluences of 92 MeV Xe ions (Se≈18 keV/nm) with a maximum flux of $2\times10^9$ ions cm$^{-2}$.s$^{-1}$ (maximum used to prevent macroscopic sample heating). Up to a fluence of $5\times10^{13}$ ions.cm$^{-2}$ no structural variation was observed, the d-spacings (i.e. line positions) change by less than 1%. The peak areas change by less than 10%, meaning that the phase remains crystalline and therefore confirming the radiation-resistant behaviour of MgO. The use of a high flux ($2\times10^9$ ions cm$^{-2}$.s$^{-1}$) of swift ions is a strong test for the discrimination of background radiation. The main problem is the emission of X-rays due to ion-target interaction, which constitutes the background to the X-ray diffraction. This background can be seen in the upper, red curve in figure 4. The energy discrimination of the Vantec detector can then be adjusted close to the $K_{\alpha1}$ Cu radiation energy used by the X-ray source for the diffraction to minimize the impact of background due to ion-target interaction. The cleaner signal with the suppressed background due to the energy discrimination can be seen in the lower, blue curve in figure 4. Due to this energy discrimination, the signal-to-noise ratio rises from 10 to 245. This experiment has shown that the detector can be used for *in-situ* X-ray diffraction experiments simultaneously with ion irradiation.

**III.   Results: *In-situ* grazing incidence X-ray diffraction of SrTiO$_3$ perovskite during xenon irradiation**

Results of 92 MeV xenon ion irradiation effects on the structural properties of polycrystalline SrTiO$_3$ and its damages created by electronic excitation are presented in this section. The structure of SrTiO$_3$ is a cubic perovskite (space group n°221: *Pm-3m*) with a lattice parameter of 3.905 Å. The polycrystalline pellets have been prepared by conventional solid state process. As shown in figure 1, the electronic stopping power diminishes strongly with increasing SrTiO$_3$ depth. If we want to study a part of the target where the energy loss changes by less than 20%, thus incident angle for X-ray beam has been fixed to 1° for a probed thickness of 700 nm and a constant $S_e$ of 20 keV/nm which is above the amorphization threshold.

Two experiments have been performed: one with *simultaneous irradiation – diffraction,* and one with *sequential irradiation – diffraction*.

For *simultaneous irradiation - diffraction* experiments, if we want to obtain enough statistics the mode does not allow to measure the full diffraction pattern of SrTiO$_3$, a 2θ range (21-42°: 2 times the detector window) is used where three main diffraction reflections are present. The target was irradiated with Xe ions with a flux of about $5\times10^7$ ions cm$^{-2}$ s$^{-1}$ for low fluence up to $5\times10^8$ ions cm$^{-2}$ s$^{-1}$ for higher fluence. A pattern is acquired every 480 s. 130 patterns are stored for fluences up to $1\times10^{14}$ ions cm$^{-2}$. The total acquisition time amounts to 18h. We used the detector energy discrimination to reduce the pattern background during ion bombardment.

A set of s*equential irradiation – diffraction* has also been performed for another SrTiO$_3$ pellet. The X-ray measurement time was 75 minutes per scan in the 10-90° 2θ range, these patterns were measured at 12 different fluences. In total, the experimental time contains 18h irradiation (until $1\times10^{14}$ ions cm$^{-2}$ at a flux of $2\times10^9$ ions cm$^{-2}$ s$^{-1}$) plus the measurement time of the patterns; the total time for this mode is therefore 40h. The energy discrimination is not necessary in this case as the patterns are acquired with the ion beam turned off.

The X-ray patterns for *sequential irradiation– diffraction* are shown in figure 5 and for *simultaneous irradiation – diffraction* in figure 6. Figure 5 shows a global decrease of the total diffracted intensity until the reflection extinctions. Simultaneously a diffuse scattering is appearing from $5\times10^{12}$ ions cm$^{-2}$ fluence at 30° 2θ value. The same observations are valid with the patterns obtained during *simultaneous irradiation – diffraction* where the reflections (only *001*, *110* and *111* are observed in the measured 2θ range) are decreasing and the diffuse scattering is growing, as seen figure 6. These results show that a phase transition from the crystalline to the amorphous state of $SrTiO_3$ is induced by swift heavy ion irradiation. In order to compare data from both modes, we are comparing the net area evolution of the (*110*) reflection (2θ = 32.5°). The amorphization fraction $F_a$ is calculated by:

$$F_a = 1 - \frac{A_{irr}}{A_v} \quad (1)$$

where $A_{irr}$ is the net area of the (*110*) reflection for the irradiated sample and $A_v$ is the net area of the virgin sample. The amorphization fraction as a function of the ion fluence is presented in figure 7a, the curves for *simultaneous irradiation - diffraction* (figure 7a) and *sequential irradiation– diffraction* are well in agreement. This shows that the simultaneous measurement is possible, allowing more detailed and faster measurements. We will use this mode in the following to distinguish between two different defect mechanisms. This comparison shows the utility of the *simultaneous irradiation - diffraction* mode.

The amorphization fraction can be modelled by the overlapping impact mechanism expressed by the Gibbons model[24]:

$$F_a = f_s \left[ 1 - \sum_{k=0}^{n-1} \frac{(\sigma_a \Phi)^k}{k!} e^{-\sigma_a \Phi} \right] \quad (2)$$

where *n* is the number of impacts necessary for the creation of defects, $f_s$ the saturation value of the amorphous fraction, $\sigma_a$ the cross-section of the amorphized cylinder and $\Phi$ the ion fluence. Figure 7b shows the fit of this function to the measured data, using either the simple impact model (*n* = 1) or the double impact model (*n* = 2). Clearly the single impact model exhibits a better agreement with the experimental data, especially at low fluences (below $1\times10^{13}$ ions cm⁻²), where the amorphization mechanism is rising and the difference between the two models is the largest. The fit of the single impact model yields a track radius of (2.4±0.4) nm, derived from $\sigma_a$. Moreover, the fit yields $f_s$ = (0.989±0.004), which means that a fully amorphous phase is obtained at high fluences.

Further observations have been done by transmission electron microscopy (TEM) with a JEOL 2010F electron microscope operating at 200kV, in order to support the XRD measurements. Figure 8 shows high resolution micrographs for a $SrTiO_3$ sample, irradiated at a low fluence ($1\times10^{12}$ ions cm⁻²) to prevent track overlapping. Figure 8a is overfocused to highlight the contrast of damaged areas and it confirms the presence of latent tracks. Contrary to the assumption of the single impact model, that the tracks are amorphous, we see with the TEM in figure 8b crystalline areas containing defects at the residual position of the latent track. This may be due to recrystallization during electron beam exposure.[15] Besides the track structure, a track radius of 2.5-3 nm has been extracted from the remaining defect area observed in high resolution images. This value corresponding well to the value of (2.4±0.4) nm, obtained from XRD.

To summarize, these experiments were performed for high energy ion irradiation on $SrTiO_3$ ceramics by using all the capabilities of the ALIX device, especially the *simultaneous irradiation – diffraction* mode using the energy discrimination option. We have shown for $SrTiO_3$ that a crystalline to amorphous transition occurs by a single ion impact mechanism, as is often used to describe the damage accumulation in irradiated materials by swift heavy ions.[25, 26]

**IV. Summary and outlook**

ALIX, a X-ray diffractometer has been set up at the IRRSUD beamline of GANIL to study structural evolutions during swift heavy ion irradiation. This setup enables simultaneous irradiation and X-ray diffraction by using energy discrimination between X-rays from diffraction and from ion-target interaction, and therefore keeping a good signal/background ratio in the patterns. On the IRRSUD beamline the energy range is from 0.3 to 1 MeV/u implying a mean ion depth penetration in solid matter around few micrometers. Therefore, grazing incidence X-ray measurements are required to measure only in the upper damaged layer, where the energy loss stays constant. This is easily possible by using a a parallel X-ray beam delivered by a Göbel mirror in the primary optics.

Phase transition kinetics of polycrystalline MgO and $SrTiO_3$ have been studied during Xe ion irradiation. We have observed that MgO oxide is radiation-resistant to high electronic excitations. This is contrary to the high irradiation sensitivity of $SrTiO_3$ which exhibits transitions from crystalline to amorphous. The amorphization kinetics of $SrTiO_3$ can be explained by a single impact model.

Finally, we have shown that ALIX allows obtaining good statistics, and therefore quick and easy access to phase transition kinetics of during swift heavy ion irradiation. Different kinds of transformations like crystal-to-amorphous, amorphous-to-crystal (ion induced recrystallization), crystal-to-crystal and nanocrystal formation or even grain reorientation to preferential orientation can now be studied. *Simultaneous irradiation – diffraction* is efficient and provides a lot of diffraction patterns at different fluence points. Only in the case of displacive or complex transitions requiring detailed analysis or Rietveld refinements,[27, 28] *sequential irradiation – diffraction* is necessary. This allows getting full X-ray patterns, unfortunately the experimental time is much longer. Actually, ALIX is the only set-up performing *simultaneous irradiation - diffraction*. It is now open to the scientific community, further details can be obtained at the website for interdisciplinary research at the GANIL facility, organised by the Centre de Recherche sur les Ions, les Matériaux et la Photonique (CIMAP).[29]

## ACKNOWLEDGMENTS

The authors want to thank David Siméone, Dominique Gosset and Guido Baldinozzi for the helpful discussion on technical definitions of the equipment. The experiments were performed at IRRSUD beamline of the Grand Accélérateur National d'Ions Lourds (GANIL), Caen, France. The authors thank the GANIL technical staff and administrative staff. This work was supported by the ANR contract ALIX-Mai ANR-06-BLAN-0292-01 and by the Region Basse-Normandie.

**Figure captions**

Figure 1: (Color online) Electronic stopping power (left-open squares) and number of displacement per atom (right-open circles, calculated for a fluence of $1\times10^{14}$ ions.cm$^{-2}$) as a function of the depth for SrTiO$_3$ irradiated by 92 MeV Xe. The hashed region shows the depth probed by X-ray diffraction at a grazing incidence of 1°.

Figure 2: ALIX diffractometer set up at IRRSUD beam line at GANIL. 1) Ion beam, 2) Turbomolecular pump, 3) Window for diffracted X-rays, 4) Window for incident X-rays and 5) X-ray generator.

Figure 3: (Color online) X-ray pattern for LaB$_6$ powder measured in conventional θ/2θ geometry and its Rietveld refinement with experimental data (Yobs, open circles), calculated data (Ycalc, black line on top) and difference between experimental and calculated data (Yobs-Ycalc, blue line on bottom, shifted for clarity).

Figure 4: (Color online) X-ray patterns without energy discrimination E$_d$ (red – right axis) and with energy discrimination (blue – left axis) for MgO polycrystalline pellet during Xe irradiation at a fluence of $3\times10^{13}$ ions cm$^{-2}$ where 100% of material remain crystalline.

Figure 5: X-ray patterns as function of Xe fluence in the mode *sequential irradiation – diffraction* for SrTiO$_3$ polycrystalline pellet.

Figure 6: X-ray patterns as function of Xe fluence in the mode *simultaneous irradiation – diffraction* for SrTiO$_3$ polycrystalline pellet. For clarity only few curves over the 130 curves are shown.

Figure 7: (Color online) a) Amorphous fraction versus Xe fluence for SrTiO$_3$ polycrystalline pellet in *simultaneous irradiation – diffraction* and in *sequential irradiation– diffraction* modes. b) Logarithmic scale showing a close up at low fluence for *simultaneous* mode fitted by single impact model (red) and by double impact model (blue).

Figure 8: High resolution images of SrTiO$_3$ polycrystalline pellet after Xe irradiation at a fluence of $1\times10^{12}$ ions cm$^{-2}$. a) The image is overfocused to highlight latent track contrasts. b) It shows recrystallization of latent track under 200kV electron beam exposure.

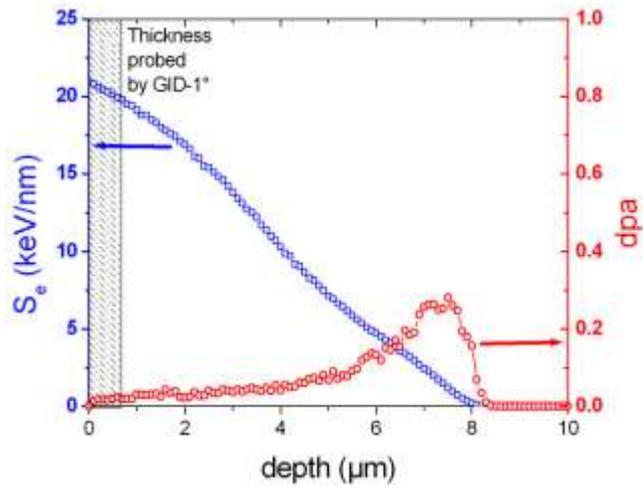

Fig. 1

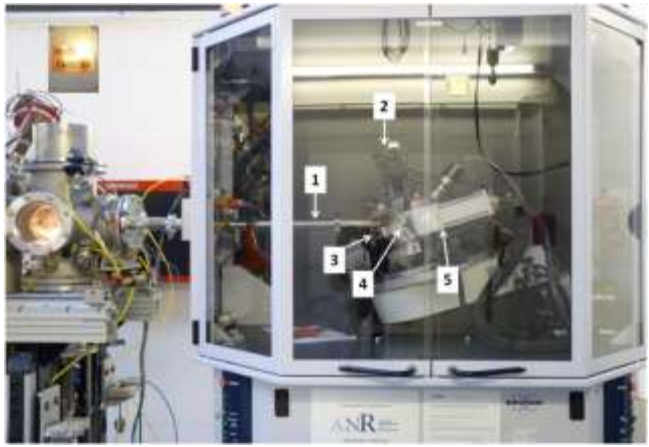

Fig. 2

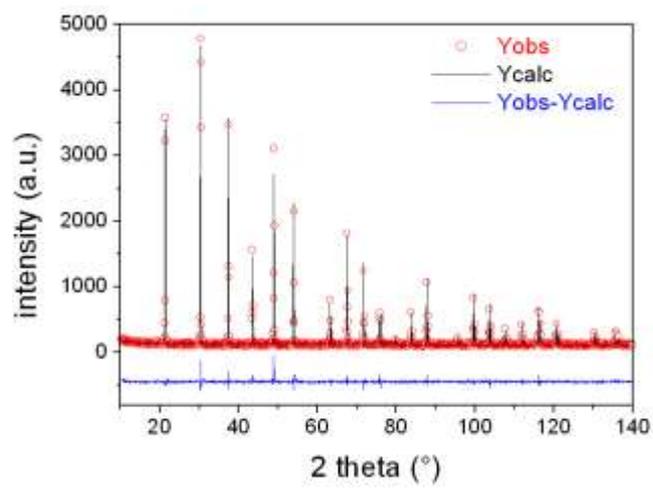

Fig. 3

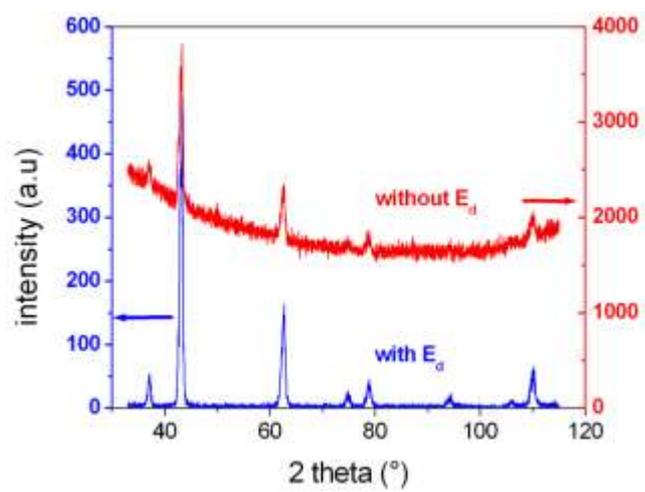

Fig. 4

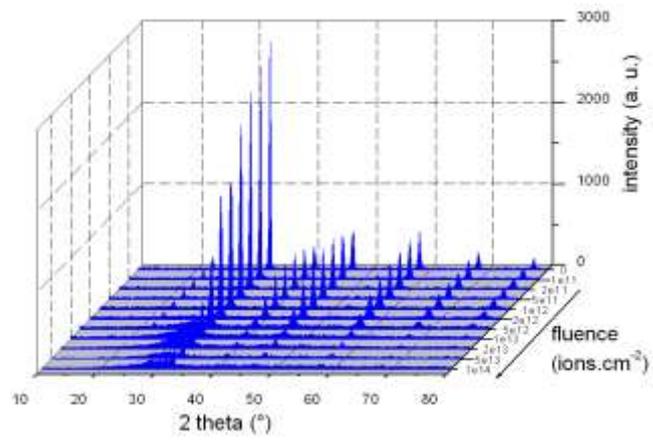

Fig. 5

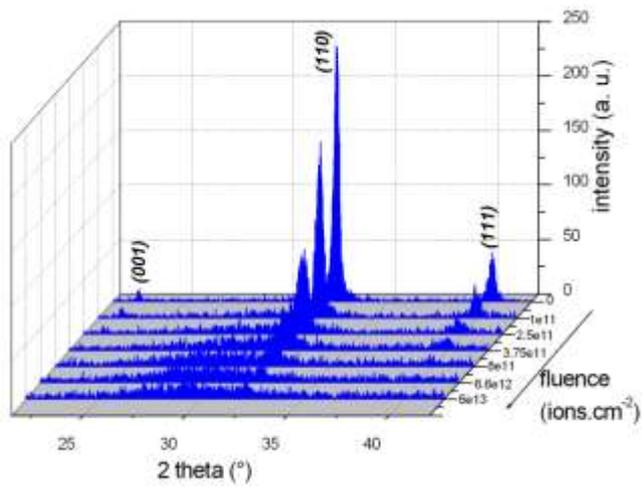

Fig. 6

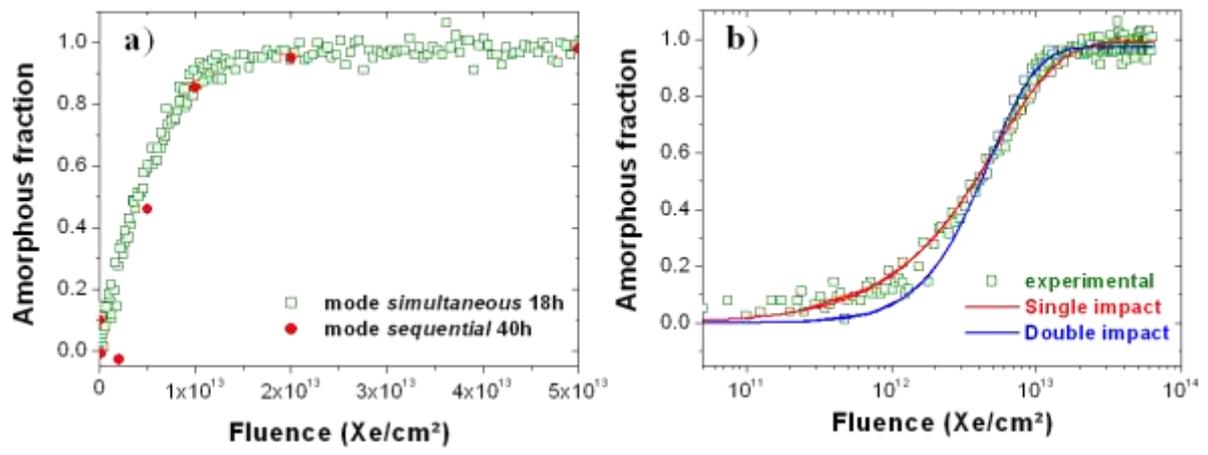

Fig. 7

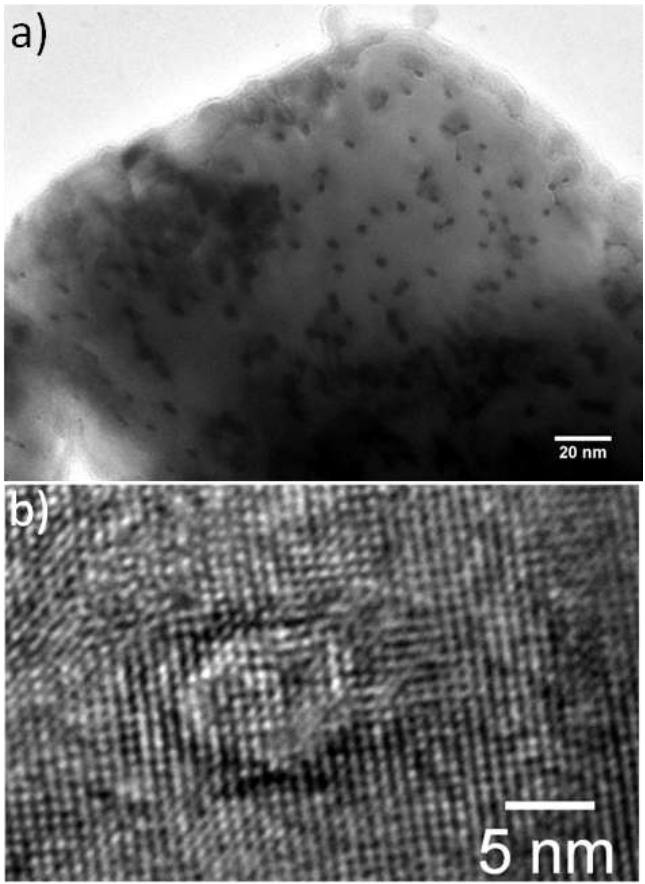

Fig. 8